\documentclass[sigconf]{acmart}
\usepackage{multirow}
\usepackage{makecell}

\AtBeginDocument{%
  \providecommand\BibTeX{{%
    \normalfont B\kern-0.5em{\scshape i\kern-0.25em b}\kern-0.8em\TeX}}}

\setcopyright{acmcopyright}
\copyrightyear{2023}
\acmYear{2023}
\acmDOI{10.1145/nnnnnnn.nnnnnnn}

\acmConference[WSDM’23]{Proceedings of the ACM
International WSDM Conference}{February,
2023}{Singapore}

\acmPrice{15.00}
\acmISBN{978-1-4503-XXXX-X/18/06}

\begin{document}

\title{Enhancing Model Performance in Multilingual Information Retrieval with Comprehensive Data Engineering Techniques}



\author{Qi Zhang, Zijian Yang, Yilun Huang, Ze Chen, Zijian Cai, Kangxu Wang, Jiewen Zheng, Jiarong He, Jin Gao}
\email{{zhangqi21,yangzijian,huangyilun,jackchen,caizijian01,wangkangxu,zhengjiewen,gzhejiarong,jgao}@corp.netease.com}
\affiliation{%
  \institution{Interactive Entertainment Group of Netease Inc.}
  \city{Guangzhou}
  \country{China}
}







\begin{abstract}
In this paper, we present our solution to the Multilingual Information Retrieval Across a Continuum of Languages (MIRACL) challenge of WSDM CUP 2023\footnote{https://project-miracl.github.io/}. Our solution focuses on enhancing the ranking stage, where we fine-tune pre-trained multilingual transformer-based models with MIRACL dataset. Our model improvement is mainly achieved through diverse data engineering techniques, including the collection of additional relevant training data, data augmentation, and negative sampling. Our fine-tuned model effectively determines the semantic relevance between queries and documents, resulting in a significant improvement in the efficiency of the multilingual information retrieval process. Finally, Our team is pleased to achieve remarkable results in this challenging competition, securing 2nd place in the Surprise-Languages track with a score of 0.835 and 3rd place in the Known-Languages track with an average nDCG@10 score of 0.716 across the 16 known languages on the final leaderboard.  
\end{abstract}



\maketitle

\section{Introduction}
The MIRACL challenge of WSDM CUP 2023 aims to assess the performance of monolingual retrieval systems across 18 diverse languages. Participants are provided with the MIRACL dataset, consisting of balanced monolingual queries and human-annotated documents. The Known-Languages track of the challenge includes 16 languages, which are evaluated using the nDCG@10 metric. The Surprise-Languages track, consisting of Deutsch and Yoruba, is also evaluated using nDCG@10. 

The task of ranking a set of textual documents based on their relevance to a query or context is known as text ranking. This is an essential component in real-world applications such as search engines and question answering systems. The recent advancements in transformer-based large language models have been remarkable and have demonstrated exceptional performance across a wide range of NLP tasks, including information retrieval, sentiment analysis, translation, question answering, and more\cite{zhang2022semantic,devlin2018bert,abolghasemi2022improving,zhang2022towards}.

In the field of information retrieval, existing solutions utilize either bag-of-words systems, such as BM25\cite{crestani1998document}, or bi-encoder models like DPR\cite{karpukhin2020dense}, to perform retrieval. The top-k candidates generated in the retrieval stage are then ranked using cross-encoder models. 

For multilingual scenarios, switching to a multilingual pre-trained language model (PLM) is sufficient. As demonstrated in the study\cite{zhang2022towards}, multilingual transformers have the ability to transfer relevance matching knowledge across languages.

Our solution builds upon the retrieval and ranking pipeline and incorporates ensemble techniques to rerank the ranking results. We use Pyserini\cite{Lin_etal_SIGIR2021_Pyserini} for the retrieval stage, with the majority of our efforts focused on improving the ranking stage. Most of our model improvement is achieved through data engineering techniques. We pre-fine-tune our models using the MS MARCO\cite{nguyen2016ms} and Mr. TyDi\cite{zhang2021mr} datasets, followed by fine-tuning with the MIRACL dataset\cite{zhang2022making}. To further enhance the models, we employ data augmentation techniques such as Query-to-Query-to-Document (Q2Q2D) and pseudo labeling, and use negative sampling techniques to address sample selection bias. The specifics of our solution will be presented in detail in the following sections.

\section{Related Work}
\subsection{Cross-Encoder Models}
The advancement of neural approaches has significantly improved the results of information retrieval in recent years. Previously, similarity metrics were largely based on keyword matching, with limited thesaurus and phrase-based expansion. The Cross-Encoder architecture has further enhanced the field of text understanding by simultaneously passing the query and product through transformer networks to produce an output representation that reflects the similarity between the input pairs\cite{reimers2019sentence}.

\subsection{Multilingual Transformer-based Models}
The challenge of defining textual features in a cross-lingual representation space becomes increasingly complex as the number of languages increases. To address this issue, XLM\cite{lample2019cross} employs Byte-Pair Encoding, which divides the input into the most frequently occurring sub-words across different languages, instead of using words or characters. Additionally, the Translation Language Modeling (TLM) task further enhances the model's ability to encode contextual information. Today, large-scale transformer-based pre-trained language models such as RemBERT\cite{chung2020rethinking}, XLM-RoBERTa\cite{conneau2019unsupervised}, InfoXLM\cite{chi2020infoxlm} and mDeBERTa\cite{he2020deberta,he2021debertav3} have set new benchmarks in various NLP tasks. It has been shown that training cross-lingual language models can lead to improved performance in many NLP applications.

\subsection{Dense Retrieval Models}
Dense Retrieval Models\cite{karpukhin2020dense,lin2021batch, hofstatter2021efficiently, xiong2020approximate} are information retrieval models based on bi-encoder architecture. Unlike traditional information retrieval models that rely on keyword matching, dense retrieval models use dense vectors to represent the complete semantic meaning of a query and document respectively. Dense retrieval models can be applied to compute document representations, leading to the creation of an index of all documents offline. Afterwards, efficient and scalable nearest neighbor searches for specific queries can be performed using existing open-source toolkits such as Faiss\cite{johnson2019billion}.

\section{Dataset Overview}
The MIRACL (Multilingual Information Retrieval Across a Continuum of Languages) dataset\cite{zhang2022making} is a unique resource for researchers working on search across multiple languages. It covers 18 different languages, each of which is divided into four parts: train, dev, test-A, and test-B. Each sample in the dataset contains a query, a passage, and a judgment indicating the relevance of the passage to the query.

Table 1 presents a comprehensive overview of the descriptive statistics of the MIRACL dataset. It includes crucial information such as the number of queries and human-annotated relevance judgments for each language-split combination. Additionally, it displays the number of passages and articles included in the corpora. With over 600k training pairs, the MIRACL dataset offers a wealth of data for researchers to work with.

In addition to the standard tracks, the MIRACL competition introduces a new and exciting challenge in the form of the Surprise-Languages track. This track includes 2 new languages that have not appeared in the training set, testing the ability of multilingual models to transfer knowledge from known languages to unseen languages.

\begin{table*}
\begin{tabular}{@{}lllllllllll@{}}
\toprule
\multirow{2}{*}{\textbf{Lang}} & \multicolumn{2}{l}{\textbf{Train}} & \multicolumn{2}{l}{\textbf{Dev}} & \multicolumn{2}{l}{\textbf{Test-A}} & \multicolumn{2}{l}{\textbf{Test-B}} & \multirow{2}{*}{\textbf{\# Passages}} & \multirow{2}{*}{\textbf{\# Articles}} \\ \cmidrule(lr){2-9}
                               & \textbf{\# Q}    & \textbf{\# J}   & \textbf{\# Q}   & \textbf{\# J}  & \textbf{\# Q}    & \textbf{\# J}    & \textbf{\# Q}    & \textbf{\# J}    &                                       &                                       \\ \cmidrule(r){1-11}
Arabic (ar)                    & 3,495            & 25,382          & 2,896           & 29,197         & 936              & 9,325            & 1,405            & 14,036           & 2,061,414                             & 656,982                               \\
Bengali (bn)                   & 1,631            & 16,754          & 411             & 4,206          & 102              & 1,037            & 1,130            & 11,286           & 297,265                               & 63,762                                \\
English (en)                   & 2,863            & 29,416          & 799             & 8,350          & 734              & 5,617            & 1,790            & 18,241           & 32,893,221                            & 5,758,285                             \\
Spanish (es)                   & 2,162            & 21,531          & 648             & 6,443          & 0                & 0                & 1,515            & 15,074           & 10,373,953                            & 1,669,181                             \\
Persian (fa)                   & 2,107            & 21,844          & 632             & 6,571          & 0                & 0                & 1,476            & 15,313           & 2,207,172                             & 857,827                               \\
Finnish (fi)                   & 2,897            & 20,350          & 1,271           & 12,008         & 1,060            & 10,586           & 711              & 7,100            & 1,883,509                             & 447,815                               \\
French (fr)                    & 1,143            & 11,426          & 343             & 3,429          & 0                & 0                & 801              & 8,008            & 14,636,953                            & 2,325,608                             \\
Hindi (hi)                     & 1169             & 11,668          & 350             & 3,494          & 0                & 0                & 819              & 8,169            & 506,264                               & 148,107                               \\
Indonesian (id)                & 4,071            & 41,358          & 960             & 9,668          & 731              & 7,430            & 611              & 6,098            & 1,446,315                             & 446,330                               \\
Japanese (ja)                  & 3,477            & 34,387          & 860             & 8,354          & 650              & 6,922            & 1,141            & 11,410           & 6,953,614                             & 1,133,444                             \\
Korean (ko)                    & 868              & 12,767          & 213             & 3,057          & 263              & 3,855            & 1,417            & 14,161           & 1,486,752                             & 437,373                               \\
Russian (ru)                   & 4,683            & 33,921          & 1,252           & 13,100         & 911              & 8,777            & 718              & 7,174            & 9,543,918                             & 1,476,045                             \\
Swahili (sw)                   & 1,901            & 9,359           & 482             & 5,092          & 638              & 6,615            & 465              & 4,620            & 131,924                               & 47,793                                \\
Telugu (te)                    & 3,452            & 18,608          & 828             & 1,606          & 594              & 5,948            & 793              & 7,920            & 518,079                               & 66,353                                \\
Thai (th)                      & 2,972            & 21,293          & 733             & 7,573          & 992              & 10,432           & 650              & 6,493            & 542,166                               & 128,179                               \\
Chinese (zh)                   & 1,312            & 13,113          & 393             & 3,928          & 0                & 0                & 920              & 9,196            & 4,934,368                             & 1,246,389                             \\ \bottomrule
\end{tabular}
\caption{Descriptive statistics of MIRACL dataset}
\end{table*}

\section{Methodology}

Our proposed solution is centered around the framework illustrated in Figure 1, which consists of three key components: retrieval, ranking, and reranking. To improve the performance of the ranking models, we have also implemented techniques such as negative sample mining and data augmentation. A more in-depth explanation of these components is provided below.

\begin{figure}
    \includegraphics[width=\linewidth]{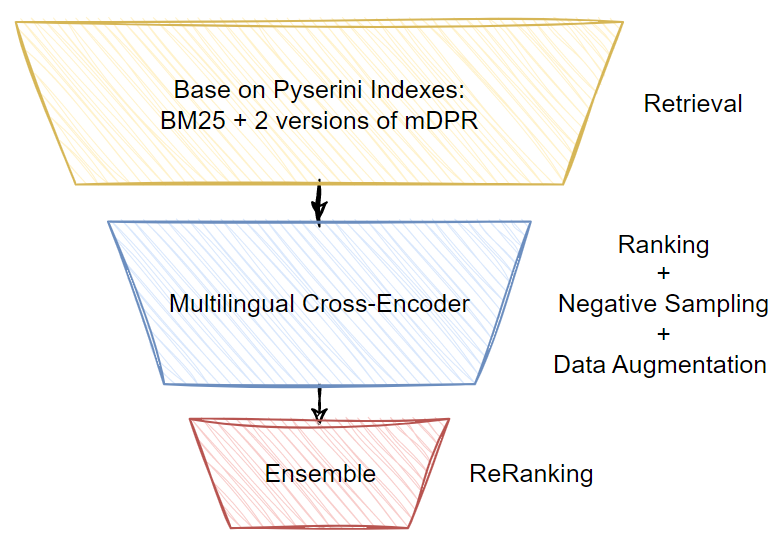}
    \caption{Framework of Our Proposed Solution}
    \label{fig:fig1}
\end{figure}

\subsection{Retrieval}

Our retrieval work is primarily reliant on Pyserini\cite{Lin_etal_SIGIR2021_Pyserini}, a user-friendly and powerful Python toolkit for information retrieval research using sparse and dense representations. Specifically, retrieval is performed in a zero-shot manner with both BM25 and mDPR retrieval indices. These methods are the baselines provided by the MIRACL Team, and the nDCG@10 scores of test-A split are 0.449 for BM25 and 0.398 for mDPR.

We then perform an ensemble of the BM25 and mDPR scores to produce a more robust retrieval result, as described in the paper\cite{zhang2022making}. The ensemble improves the score to 0.635 on the public leaderboard, and the top 200 candidates from the ensemble result are selected for subsequent ranking.

The annotation workflow of MIRACL dataset is illustrated in Figure 2. Relevance assessment was done by the annotators based on the ensemble results of 3 different retrieval methods, which are BM25, mDPR and ColBERT.

\begin{figure}
    \includegraphics[width=\linewidth]{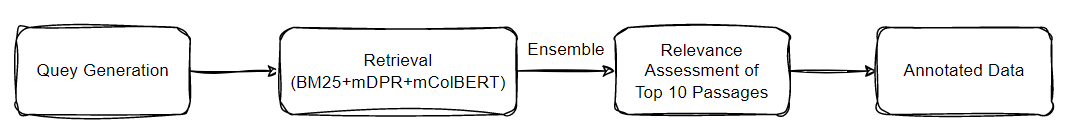}
    \caption{Annotation Workflow for the MIRACL Dataset}
    \label{fig:fig2}
\end{figure}

Table 2 presents the top 200 recall rates for the languages in the training set. The majority of the results range from 0.980 to 0.999, with only two exceptions: English (0.9646) and Indonesian (0.9522). Although there is a slight decrease in recall, likely due to missed information from ColBERT as noted in the previous study\cite{zhang2022making}. We believe that the top 200 recall rate for the combined BM25 and mDPR results in the training set is sufficient for ranking models. Given these results, we have not made additional improvements to the retrieval models. Instead, we have utilized the top 200 retrieval results of the BM25 and mDPR hybrid for the next step of ranking.

\begin{table*}
  
  \begin{tabular}{ccccccccccccccccc}
    \toprule
    Language & ar & bn & en & es & fa & fi & fr & hi & id & ja & ko & ru & sw & te & th & zh\\
    \midrule
    recall\_rate &0.993 & 0.999 & 0.965 & 0.991 & 0.980 & 0.995 & 0.995 & 0.999 & 0.952 & 0.995 & 0.984 & 0.984 & 0.990 & 0.988 & 0.996 & 0.996 \\
    \bottomrule
  \end{tabular}
  \caption{Top 200 recall rates for the languages in the training set}
\end{table*}

\subsection{Ranking}

After conducting retrieval, we employ the MS MARCO passage dataset for pre-fine-tuning, as suggested in the study\cite{zhang2022towards}. Subsequently, we utilize the MIRACL annotation data for further refinement of the ranking process. Our ranking model is built on a multilingual cross-encoder architecture, and we fine-tune it using 3 different PLMs for ensemble predictions. These PLMs include RemBERT, InfoXLM and mDeBERTa. During preprocessing, the texts of the query, title, and document are concatenated and tokenized. To ensure optimal input, the resulting text is truncated to a maximum length of 256. The [CLS] embedding is then utilized for the downstream task of binary classification.

Adversarial training is an effective technique that enhances the robustness of neural networks by making them more resistant to adversarial attacks. We employ Fast Gradient Method (FGM)\cite{goodfellow2014explaining} for adversarial training and are able to achieve an improvement of 0.003 on the leaderboard score. 

Dropout is a powerful regularization technique that helps prevent overfitting and achieve better generalization. By combining multiple dropout layers with varying dropout ratios, multi-sample dropout\cite{inoue2019multi} provides even greater improvement for the model. In our scenario, we utilize multi-sample dropout before the output layer to make our model more robust. To further increase efficiency during training, we employ mixed precision. Additionally, we utilize negative sampling and data augmentation to enhance the model's performance.

\subsection{Negative Sampling}

As discussed in the MIRACL paper, the annotation is based on the top 10 passages generated by an ensemble of separate retrieval models. Traditional supervised machine learning assumes that the training and test data are independently and identically distributed (i.i.d.). Using the annotation data directly for training the ranking model may result in a sample selection bias\cite{marlin2012collaborative, schnabel2016recommendations}, which can negatively impact the final performance of the model. To address this issue, we add additional  negative samples to the annotation data, which are randomly selected from the top 200 ensemble retrieval results. As the number of negative samples increases, the model performance improves smoothly. This method is found to be more effective than using negative samples from the entire document corpus.

\subsection{Data Augmentation}
In order to enhance the performance of the ranking model, we employ two methods of data augmentation: Q2Q2D and pseudo labeling. These techniques provide additional training data, helping the model to generalize better and produce more accurate results.

\subsubsection{Q2Q2D}
The basic concept behind Query-to-Query-to-Document (Q2Q2D) is to link similar queries from the test set to those in the train set and use the selected train set queries and their associated annotation data for data augmentation. \\
We utilize sentence transformers with the pretrained model weights \verb|paraphrase-multilingual-mpnet-base-v2|\cite{bharathi2021machine} to calculate the similarity between the test and train set queries. The similarity scores are then multiplied by the annotated labels and a parameter called Alpha to generate the final label for the augmented data. Alpha is used to control any potential noise introduced and is set to 0.9 in most of our experiments.

\begin{figure}
    \includegraphics[width=\linewidth]{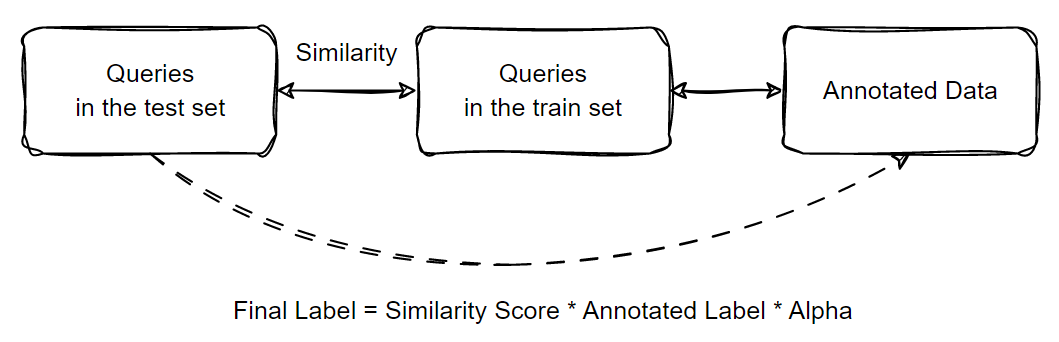}
    \caption{The Q2Q2D Data Augmentation Workflow}
    \label{fig:fig3}
\end{figure}

\subsubsection{Pseudo Labeling}

In addition to Q2Q2D, we also utilize pseudo labeling\cite{lee2013pseudo} as another method of data augmentation. By using our trained models to generate pseudo labels from the test set, we are able to increase the amount of data used for training and improve the model's adaptation to the test set queries and documents. In our experiments, soft labels outperform hard labels, as the latter may increase the risk of overfitting. To maintain the quality of the training data, we randomly select a portion of the pseudo labeling data and multiply the soft label by 0.9 to produce the final label used for training. This technique brings a significant improvement to our models.

\subsection{Reranking}

Finally, we utilize model ensemble to further improve the ranking results and achieve an improvement of 0.008 on the public leaderboard score. Our base models for ensemble include RemBERT, InfoXLM, and mDeBERTa, each trained with various settings as described above. The weights for averaging the predictions of these models are primarily determined by their scores on the public leaderboard. Additionally, we decrease the weights of models with high correlation coefficients in their prediction results to ensure a diverse ensemble.

\section{Results}

\begin{table*}
\begin{tabular}{@{}lll@{}}
    \toprule
Methods                                     & \makecell[l]{Known-Languages track \\ Test-A Split \\ nDCG@10 [Private Score]} & \makecell[l]{Surprise-Languages track \\ Dev Split \\ nDCG@10 [Private Score]} \\
    \midrule
MIRACL (mDPR)                               & 0.398                                          & 0.467                                          \\
MIRACL (BM25)                               & 0.449                                          & 0.316                                          \\
mDPR+BM25 Hybrid                            & 0.635                                          & -                                              \\
\hline
mDeBERTa Ranking Baseline                   & 0.701                                          & -                                              \\
\hline
\textbf{mDeBERTa PFT with MS MARCO (PFT )}  & 0.730                                          & -                                              \\
mDeBERTa + PFT + PL                         & 0.763                                          & -                                              \\
mDeBERTa + PFT (+ Mr. TyDi) + PL                  & 0.768                                          & -                                              \\
mDeBERTa + PFT (+ Mr. TyDi) + PL + Q2Q2D          & 0.778                                          & -                                              \\
mDeBERTa + PFT (+ Mr. TyDi) + PL + Q2Q2D + NS-100 & 0.785                                          & -                                              \\
\hline
\textbf{RemBERT + PFT}                      & 0.744                                          & -                                              \\
RemBERT + PFT + PL                          & 0.772                                          & -                                              \\
RemBERT + PFT + PL + NS-10                  & 0.781                                          & -                                              \\
RemBERT + PFT + PL + NS-100                 & 0.792                                          & 0.839                                          \\
\hline
\textbf{InfoXLM-large + PFT}                & 0.745                                          & -                                              \\
InfoXLM-large + PFT (+ Mr. TyDi) + NS-5           & 0.764                                          & -                                              \\
InfoXLM-large + PFT (+ Mr. TyDi) + NS-50          & 0.771                                          & -                                              \\
InfoXLM-large + PFT (+ Mr. TyDi) + NS-100         & 0.799                                          & 0.843                                          \\
InfoXLM-large + PFT (+ Mr. TyDi) + NS-200         & 0.802                                          & -                                              \\
\hline
\textbf{Ensemble}                           & \textbf{0.810 [0.716]}                                          & \textbf{0.859 [0.835]}           \\
    \bottomrule
\end{tabular}
\caption{Results of Experiments}
\end{table*}

The results of our experiments on the public leaderboard are presented in Table 3. Our solutions are consistent across both the Known-Languages track and the Surprise-Languages track.

Our experiment is based on the official retrieval baseline, which utilizes BM25 and mDPR. The nDCG@10 scores for BM25 and mDPR on the Test-A split are 0.449 and 0.398, respectively. By averaging the results of BM25 and mDPR, the score increases to 0.635. We make minimal adjustments to the retrieval stage and simply use the top 200 ensemble results of BM25 and mDPR for subsequent ranking and reranking.

All of our ranking models are based on cross-encoder architectures. We start with mDberta as our baseline model with a classification head, which yields a score of 0.701 on the leaderboard. Further improvements are made through pre-fine-tuning (PFT) with additional data sources such as MS MARCO and Mr. TyDi, negative sampling, and data augmentation techniques such as pseudo labeling and Q2Q2D. PFT with MS MARCO is particularly effective. Building on this, we utilize the multilingual version of MS MARCO for further pre-fine-tuning. However, the improvement is not as substantial as we hope. Our models, RemBERT and InfoXLM-large, perform better than mDeBERTa with scores of 0.744 and 0.745, respectively. Although mDeBERTa's score of 0.730 is lower, it still contributes positively to the ensemble. Pseudo labeling brings a significant improvement of approximately 0.03 to the nDCG@10 scores on the public leaderboard, however, we realize it probably causes overfitting to some extent when the private leaderboard is open. This requires further investigation. Q2Q2D is also highly effective, improving mDeBERTa's score from 0.768 to 0.778.

Initially, our negative sampling strategy involved using random samples from the entire document corpus. However, the improvement was limited. Near the end of the competition, we discovered a sample selection bias that was hindering model performance. To address this, we add random negative samples from the top 200 ensemble retrieval results in the training set that are not annotated as positive. The model shows consistent improvement as the number of negative samples increases. As shown in Table 2, InfoXLM-large's score improves from 0.764 to 0.802 as the number of negative samples increases from 5 to 200. We find that 100 negative samples are a good trade-off, as the improvement is much more pronounced when increasing the number of negative samples from 5 to 100 than from 100 to 200.

However, using pseudo labeling and Q2Q2D in conjunction with negative sampling proves challenging, as limited improvement is achieved when using all three methods simultaneously. After thorough experimentation, our best solo model achieves a score of 0.802 on the public leaderboard. This model is built using an InfoXLM-large architecture and is pre-fine-tuned with both MS MARCO and Mr. TyDi data. Additionally, it is further fine-tuned with MIRACL data and used 200 negative samples generated from the ensemble retrieval results.

With a simple ensemble, we achieve the following scores on the public leaderboard: 0.810 for the Known-Languages track, ranked 2nd, and 0.859 for the Surprise-Languages track, ranked 3rd. On the private leaderboard, our scores are 0.716 for the Known-Languages track, ranked 3rd, and 0.835 for the Surprise-Languages track, ranked 2nd.

\section{Conclusion}

In this paper, we present a comprehensive solution to the Multilingual Information Retrieval Across a Continuum of Languages (MIRACL) challenge of the WSDM CUP 2023. Our solution framework comprises three main components: retrieval, ranking, and reranking. Our focus lies in the ranking stage, where we carry out a thorough investigation. Initially, we start our experiments with multilingual cross-encoder models. To further enhance the model's performance, we first pre-fine-tune the ranking model with more relevant data, such as MS MARCO and Mr. TyDi, prior to fine-tuning with the MIRACL dataset. Given that pre-fine-tuning with more data results in a substantial improvement, we believe that data augmentation would also be beneficial. Indeed, the use of pseudo labeling and Q2Q2D techniques further improve the performance of our ranking model. Finally, to address the issue of sample selection bias, we incorporate random negative samples from the top 200 ensemble retrieval results of the training set, which results in a smooth improvement as the number of negative samples increase. Finally, we perform model ensemble to achieve the final improvement.

\begin{acks}
We extend our heartfelt gratitude to the organizing team of the MIRACL competition for their diligent efforts and contributions throughout the competition. Their tireless work in hosting such a great competition and providing high-quality datasets is greatly appreciated. We would also like to extend our thanks to everyone involved in organizing and sponsoring the WSDM 2023. Their support and commitment have made this competition a huge success.
\end{acks}


\bibliographystyle{ACM-Reference-Format}
\bibliography{sample-base}


\begin{thebibliography}{27}


\ifx \showCODEN    \undefined \def \showCODEN     #1{\unskip}     \fi
\ifx \showDOI      \undefined \def \showDOI       #1{#1}\fi
\ifx \showISBNx    \undefined \def \showISBNx     #1{\unskip}     \fi
\ifx \showISBNxiii \undefined \def \showISBNxiii  #1{\unskip}     \fi
\ifx \showISSN     \undefined \def \showISSN      #1{\unskip}     \fi
\ifx \showLCCN     \undefined \def \showLCCN      #1{\unskip}     \fi
\ifx \shownote     \undefined \def \shownote      #1{#1}          \fi
\ifx \showarticletitle \undefined \def \showarticletitle #1{#1}   \fi
\ifx \showURL      \undefined \def \showURL       {\relax}        \fi
\providecommand\bibfield[2]{#2}
\providecommand\bibinfo[2]{#2}
\providecommand\natexlab[1]{#1}
\providecommand\showeprint[2][]{arXiv:#2}

\bibitem[Abolghasemi et~al\mbox{.}(2022)]%
        {abolghasemi2022improving}
\bibfield{author}{\bibinfo{person}{Amin Abolghasemi}, \bibinfo{person}{Suzan
  Verberne}, {and} \bibinfo{person}{Leif Azzopardi}.}
  \bibinfo{year}{2022}\natexlab{}.
\newblock \showarticletitle{Improving BERT-based query-by-document retrieval
  with multi-task optimization}. In \bibinfo{booktitle}{\emph{European
  Conference on Information Retrieval}}. Springer, \bibinfo{pages}{3--12}.
\newblock


\bibitem[Bharathi and Samyuktha(2021)]%
        {bharathi2021machine}
\bibfield{author}{\bibinfo{person}{B Bharathi} {and} \bibinfo{person}{GU
  Samyuktha}.} \bibinfo{year}{2021}\natexlab{}.
\newblock \showarticletitle{Machine learning based approach for sentiment
  Analysis on Multilingual Code Mixing Text}. In
  \bibinfo{booktitle}{\emph{Working Notes of FIRE 2021-Forum for Information
  Retrieval Evaluation (Online). CEUR}}.
\newblock


\bibitem[Chi et~al\mbox{.}(2020)]%
        {chi2020infoxlm}
\bibfield{author}{\bibinfo{person}{Zewen Chi}, \bibinfo{person}{Li Dong},
  \bibinfo{person}{Furu Wei}, \bibinfo{person}{Nan Yang},
  \bibinfo{person}{Saksham Singhal}, \bibinfo{person}{Wenhui Wang},
  \bibinfo{person}{Xia Song}, \bibinfo{person}{Xian-Ling Mao},
  \bibinfo{person}{Heyan Huang}, {and} \bibinfo{person}{Ming Zhou}.}
  \bibinfo{year}{2020}\natexlab{}.
\newblock \showarticletitle{InfoXLM: An information-theoretic framework for
  cross-lingual language model pre-training}.
\newblock \bibinfo{journal}{\emph{arXiv preprint arXiv:2007.07834}}
  (\bibinfo{year}{2020}).
\newblock


\bibitem[Chung et~al\mbox{.}(2020)]%
        {chung2020rethinking}
\bibfield{author}{\bibinfo{person}{Hyung~Won Chung}, \bibinfo{person}{Thibault
  Fevry}, \bibinfo{person}{Henry Tsai}, \bibinfo{person}{Melvin Johnson}, {and}
  \bibinfo{person}{Sebastian Ruder}.} \bibinfo{year}{2020}\natexlab{}.
\newblock \showarticletitle{Rethinking embedding coupling in pre-trained
  language models}.
\newblock \bibinfo{journal}{\emph{arXiv preprint arXiv:2010.12821}}
  (\bibinfo{year}{2020}).
\newblock


\bibitem[Conneau et~al\mbox{.}(2019)]%
        {conneau2019unsupervised}
\bibfield{author}{\bibinfo{person}{Alexis Conneau}, \bibinfo{person}{Kartikay
  Khandelwal}, \bibinfo{person}{Naman Goyal}, \bibinfo{person}{Vishrav
  Chaudhary}, \bibinfo{person}{Guillaume Wenzek}, \bibinfo{person}{Francisco
  Guzm{\'a}n}, \bibinfo{person}{Edouard Grave}, \bibinfo{person}{Myle Ott},
  \bibinfo{person}{Luke Zettlemoyer}, {and} \bibinfo{person}{Veselin
  Stoyanov}.} \bibinfo{year}{2019}\natexlab{}.
\newblock \showarticletitle{Unsupervised cross-lingual representation learning
  at scale}.
\newblock \bibinfo{journal}{\emph{arXiv preprint arXiv:1911.02116}}
  (\bibinfo{year}{2019}).
\newblock


\bibitem[Crestani et~al\mbox{.}(1998)]%
        {crestani1998document}
\bibfield{author}{\bibinfo{person}{Fabio Crestani}, \bibinfo{person}{Mounia
  Lalmas}, \bibinfo{person}{Cornelis~J Van~Rijsbergen}, {and}
  \bibinfo{person}{Iain Campbell}.} \bibinfo{year}{1998}\natexlab{}.
\newblock \showarticletitle{“Is this document relevant?… probably” a
  survey of probabilistic models in information retrieval}.
\newblock \bibinfo{journal}{\emph{ACM Computing Surveys (CSUR)}}
  \bibinfo{volume}{30}, \bibinfo{number}{4} (\bibinfo{year}{1998}),
  \bibinfo{pages}{528--552}.
\newblock


\bibitem[Devlin et~al\mbox{.}(2018)]%
        {devlin2018bert}
\bibfield{author}{\bibinfo{person}{Jacob Devlin}, \bibinfo{person}{Ming-Wei
  Chang}, \bibinfo{person}{Kenton Lee}, {and} \bibinfo{person}{Kristina
  Toutanova}.} \bibinfo{year}{2018}\natexlab{}.
\newblock \showarticletitle{Bert: Pre-training of deep bidirectional
  transformers for language understanding}.
\newblock \bibinfo{journal}{\emph{arXiv preprint arXiv:1810.04805}}
  (\bibinfo{year}{2018}).
\newblock


\bibitem[Goodfellow et~al\mbox{.}(2014)]%
        {goodfellow2014explaining}
\bibfield{author}{\bibinfo{person}{Ian~J Goodfellow}, \bibinfo{person}{Jonathon
  Shlens}, {and} \bibinfo{person}{Christian Szegedy}.}
  \bibinfo{year}{2014}\natexlab{}.
\newblock \showarticletitle{Explaining and harnessing adversarial examples}.
\newblock \bibinfo{journal}{\emph{arXiv preprint arXiv:1412.6572}}
  (\bibinfo{year}{2014}).
\newblock


\bibitem[He et~al\mbox{.}(2021)]%
        {he2021debertav3}
\bibfield{author}{\bibinfo{person}{Pengcheng He}, \bibinfo{person}{Jianfeng
  Gao}, {and} \bibinfo{person}{Weizhu Chen}.} \bibinfo{year}{2021}\natexlab{}.
\newblock \showarticletitle{Debertav3: Improving deberta using electra-style
  pre-training with gradient-disentangled embedding sharing}.
\newblock \bibinfo{journal}{\emph{arXiv preprint arXiv:2111.09543}}
  (\bibinfo{year}{2021}).
\newblock


\bibitem[He et~al\mbox{.}(2020)]%
        {he2020deberta}
\bibfield{author}{\bibinfo{person}{Pengcheng He}, \bibinfo{person}{Xiaodong
  Liu}, \bibinfo{person}{Jianfeng Gao}, {and} \bibinfo{person}{Weizhu Chen}.}
  \bibinfo{year}{2020}\natexlab{}.
\newblock \showarticletitle{Deberta: Decoding-enhanced bert with disentangled
  attention}.
\newblock \bibinfo{journal}{\emph{arXiv preprint arXiv:2006.03654}}
  (\bibinfo{year}{2020}).
\newblock


\bibitem[Hofst{\"a}tter et~al\mbox{.}(2021)]%
        {hofstatter2021efficiently}
\bibfield{author}{\bibinfo{person}{Sebastian Hofst{\"a}tter},
  \bibinfo{person}{Sheng-Chieh Lin}, \bibinfo{person}{Jheng-Hong Yang},
  \bibinfo{person}{Jimmy Lin}, {and} \bibinfo{person}{Allan Hanbury}.}
  \bibinfo{year}{2021}\natexlab{}.
\newblock \showarticletitle{Efficiently teaching an effective dense retriever
  with balanced topic aware sampling}. In \bibinfo{booktitle}{\emph{Proceedings
  of the 44th International ACM SIGIR Conference on Research and Development in
  Information Retrieval}}. \bibinfo{pages}{113--122}.
\newblock


\bibitem[Inoue(2019)]%
        {inoue2019multi}
\bibfield{author}{\bibinfo{person}{Hiroshi Inoue}.}
  \bibinfo{year}{2019}\natexlab{}.
\newblock \showarticletitle{Multi-sample dropout for accelerated training and
  better generalization}.
\newblock \bibinfo{journal}{\emph{arXiv preprint arXiv:1905.09788}}
  (\bibinfo{year}{2019}).
\newblock


\bibitem[Johnson et~al\mbox{.}(2019)]%
        {johnson2019billion}
\bibfield{author}{\bibinfo{person}{Jeff Johnson}, \bibinfo{person}{Matthijs
  Douze}, {and} \bibinfo{person}{Herv{\'e} J{\'e}gou}.}
  \bibinfo{year}{2019}\natexlab{}.
\newblock \showarticletitle{Billion-scale similarity search with gpus}.
\newblock \bibinfo{journal}{\emph{IEEE Transactions on Big Data}}
  \bibinfo{volume}{7}, \bibinfo{number}{3} (\bibinfo{year}{2019}),
  \bibinfo{pages}{535--547}.
\newblock


\bibitem[Karpukhin et~al\mbox{.}(2020)]%
        {karpukhin2020dense}
\bibfield{author}{\bibinfo{person}{Vladimir Karpukhin}, \bibinfo{person}{Barlas
  O{\u{g}}uz}, \bibinfo{person}{Sewon Min}, \bibinfo{person}{Patrick Lewis},
  \bibinfo{person}{Ledell Wu}, \bibinfo{person}{Sergey Edunov},
  \bibinfo{person}{Danqi Chen}, {and} \bibinfo{person}{Wen-tau Yih}.}
  \bibinfo{year}{2020}\natexlab{}.
\newblock \showarticletitle{Dense passage retrieval for open-domain question
  answering}.
\newblock \bibinfo{journal}{\emph{arXiv preprint arXiv:2004.04906}}
  (\bibinfo{year}{2020}).
\newblock


\bibitem[Lample and Conneau(2019)]%
        {lample2019cross}
\bibfield{author}{\bibinfo{person}{Guillaume Lample} {and}
  \bibinfo{person}{Alexis Conneau}.} \bibinfo{year}{2019}\natexlab{}.
\newblock \showarticletitle{Cross-lingual language model pretraining}.
\newblock \bibinfo{journal}{\emph{arXiv preprint arXiv:1901.07291}}
  (\bibinfo{year}{2019}).
\newblock


\bibitem[Lee et~al\mbox{.}(2013)]%
        {lee2013pseudo}
\bibfield{author}{\bibinfo{person}{Dong-Hyun Lee} {et~al\mbox{.}}}
  \bibinfo{year}{2013}\natexlab{}.
\newblock \showarticletitle{Pseudo-label: The simple and efficient
  semi-supervised learning method for deep neural networks}. In
  \bibinfo{booktitle}{\emph{Workshop on challenges in representation learning,
  ICML}}, Vol.~\bibinfo{volume}{3}. \bibinfo{pages}{896}.
\newblock


\bibitem[Lin et~al\mbox{.}(2021a)]%
        {Lin_etal_SIGIR2021_Pyserini}
\bibfield{author}{\bibinfo{person}{Jimmy Lin}, \bibinfo{person}{Xueguang Ma},
  \bibinfo{person}{Sheng-Chieh Lin}, \bibinfo{person}{Jheng-Hong Yang},
  \bibinfo{person}{Ronak Pradeep}, {and} \bibinfo{person}{Rodrigo Nogueira}.}
  \bibinfo{year}{2021}\natexlab{a}.
\newblock \showarticletitle{{Pyserini}: A {Python} Toolkit for Reproducible
  Information Retrieval Research with Sparse and Dense Representations}. In
  \bibinfo{booktitle}{\emph{Proceedings of the 44th Annual International ACM
  SIGIR Conference on Research and Development in Information Retrieval (SIGIR
  2021)}}. \bibinfo{pages}{2356--2362}.
\newblock


\bibitem[Lin et~al\mbox{.}(2021b)]%
        {lin2021batch}
\bibfield{author}{\bibinfo{person}{Sheng-Chieh Lin},
  \bibinfo{person}{Jheng-Hong Yang}, {and} \bibinfo{person}{Jimmy Lin}.}
  \bibinfo{year}{2021}\natexlab{b}.
\newblock \showarticletitle{In-batch negatives for knowledge distillation with
  tightly-coupled teachers for dense retrieval}. In
  \bibinfo{booktitle}{\emph{Proceedings of the 6th Workshop on Representation
  Learning for NLP (RepL4NLP-2021)}}. \bibinfo{pages}{163--173}.
\newblock


\bibitem[Marlin et~al\mbox{.}(2012)]%
        {marlin2012collaborative}
\bibfield{author}{\bibinfo{person}{Benjamin Marlin}, \bibinfo{person}{Richard~S
  Zemel}, \bibinfo{person}{Sam Roweis}, {and} \bibinfo{person}{Malcolm
  Slaney}.} \bibinfo{year}{2012}\natexlab{}.
\newblock \showarticletitle{Collaborative filtering and the missing at random
  assumption}.
\newblock \bibinfo{journal}{\emph{arXiv preprint arXiv:1206.5267}}
  (\bibinfo{year}{2012}).
\newblock


\bibitem[Nguyen et~al\mbox{.}(2016)]%
        {nguyen2016ms}
\bibfield{author}{\bibinfo{person}{Tri Nguyen}, \bibinfo{person}{Mir
  Rosenberg}, \bibinfo{person}{Xia Song}, \bibinfo{person}{Jianfeng Gao},
  \bibinfo{person}{Saurabh Tiwary}, \bibinfo{person}{Rangan Majumder}, {and}
  \bibinfo{person}{Li Deng}.} \bibinfo{year}{2016}\natexlab{}.
\newblock \showarticletitle{MS MARCO: A human generated machine reading
  comprehension dataset}.
\newblock \bibinfo{journal}{\emph{choice}}  \bibinfo{volume}{2640}
  (\bibinfo{year}{2016}), \bibinfo{pages}{660}.
\newblock


\bibitem[Reimers and Gurevych(2019)]%
        {reimers2019sentence}
\bibfield{author}{\bibinfo{person}{Nils Reimers} {and} \bibinfo{person}{Iryna
  Gurevych}.} \bibinfo{year}{2019}\natexlab{}.
\newblock \showarticletitle{Sentence-bert: Sentence embeddings using siamese
  bert-networks}.
\newblock \bibinfo{journal}{\emph{arXiv preprint arXiv:1908.10084}}
  (\bibinfo{year}{2019}).
\newblock


\bibitem[Schnabel et~al\mbox{.}(2016)]%
        {schnabel2016recommendations}
\bibfield{author}{\bibinfo{person}{Tobias Schnabel}, \bibinfo{person}{Adith
  Swaminathan}, \bibinfo{person}{Ashudeep Singh}, \bibinfo{person}{Navin
  Chandak}, {and} \bibinfo{person}{Thorsten Joachims}.}
  \bibinfo{year}{2016}\natexlab{}.
\newblock \showarticletitle{Recommendations as treatments: Debiasing learning
  and evaluation}. In \bibinfo{booktitle}{\emph{international conference on
  machine learning}}. PMLR, \bibinfo{pages}{1670--1679}.
\newblock


\bibitem[Xiong et~al\mbox{.}(2020)]%
        {xiong2020approximate}
\bibfield{author}{\bibinfo{person}{Lee Xiong}, \bibinfo{person}{Chenyan Xiong},
  \bibinfo{person}{Ye Li}, \bibinfo{person}{Kwok-Fung Tang},
  \bibinfo{person}{Jialin Liu}, \bibinfo{person}{Paul Bennett},
  \bibinfo{person}{Junaid Ahmed}, {and} \bibinfo{person}{Arnold Overwijk}.}
  \bibinfo{year}{2020}\natexlab{}.
\newblock \showarticletitle{Approximate nearest neighbor negative contrastive
  learning for dense text retrieval}.
\newblock \bibinfo{journal}{\emph{arXiv preprint arXiv:2007.00808}}
  (\bibinfo{year}{2020}).
\newblock


\bibitem[Zhang et~al\mbox{.}(2022c)]%
        {zhang2022semantic}
\bibfield{author}{\bibinfo{person}{Qi Zhang}, \bibinfo{person}{Zijian Yang},
  \bibinfo{person}{Yilun Huang}, \bibinfo{person}{Ze Chen},
  \bibinfo{person}{Zijian Cai}, \bibinfo{person}{Kangxu Wang},
  \bibinfo{person}{Jiewen Zheng}, \bibinfo{person}{Jiarong He}, {and}
  \bibinfo{person}{Jin Gao}.} \bibinfo{year}{2022}\natexlab{c}.
\newblock \showarticletitle{A Semantic Alignment System for Multilingual
  Query-Product Retrieval}.
\newblock \bibinfo{journal}{\emph{arXiv preprint arXiv:2208.02958}}
  (\bibinfo{year}{2022}).
\newblock


\bibitem[Zhang et~al\mbox{.}(2021)]%
        {zhang2021mr}
\bibfield{author}{\bibinfo{person}{Xinyu Zhang}, \bibinfo{person}{Xueguang Ma},
  \bibinfo{person}{Peng Shi}, {and} \bibinfo{person}{Jimmy Lin}.}
  \bibinfo{year}{2021}\natexlab{}.
\newblock \showarticletitle{Mr. TyDi: A multi-lingual benchmark for dense
  retrieval}.
\newblock \bibinfo{journal}{\emph{arXiv preprint arXiv:2108.08787}}
  (\bibinfo{year}{2021}).
\newblock


\bibitem[Zhang et~al\mbox{.}(2022a)]%
        {zhang2022towards}
\bibfield{author}{\bibinfo{person}{Xinyu Zhang}, \bibinfo{person}{Kelechi
  Ogueji}, \bibinfo{person}{Xueguang Ma}, {and} \bibinfo{person}{Jimmy Lin}.}
  \bibinfo{year}{2022}\natexlab{a}.
\newblock \showarticletitle{Towards best practices for training multilingual
  dense retrieval models}.
\newblock \bibinfo{journal}{\emph{arXiv preprint arXiv:2204.02363}}
  (\bibinfo{year}{2022}).
\newblock


\bibitem[Zhang et~al\mbox{.}(2022b)]%
        {zhang2022making}
\bibfield{author}{\bibinfo{person}{Xinyu Zhang}, \bibinfo{person}{Nandan
  Thakur}, \bibinfo{person}{Odunayo Ogundepo}, \bibinfo{person}{Ehsan
  Kamalloo}, \bibinfo{person}{David Alfonso-Hermelo},
  \bibinfo{person}{Xiaoguang Li}, \bibinfo{person}{Qun Liu},
  \bibinfo{person}{Mehdi Rezagholizadeh}, {and} \bibinfo{person}{Jimmy Lin}.}
  \bibinfo{year}{2022}\natexlab{b}.
\newblock \showarticletitle{Making a MIRACL: Multilingual Information Retrieval
  Across a Continuum of Languages}.
\newblock \bibinfo{journal}{\emph{arXiv preprint arXiv:2210.09984}}
  (\bibinfo{year}{2022}).
\newblock


\end{thebibliography}










\end{document}